\documentclass[prl,showpacs,twocolumn,superscriptaddress]{revtex4}
\usepackage{bm}
\usepackage{amsfonts}
\begin{document}
\def\Barcelo{Barcel\'o}
\def\Schrodinger{Schr\"odinger}
\title[Braneworld physics from ...\dots]{Braneworld physics from the 
analog-gravity perspective: finiteness effects}
\author{Carlos \Barcelo}
\affiliation{Institute of Cosmology and Gravitation, 
University of Portsmouth, Portsmouth PO1 2EG, Britain}
\author{Antonio Campos}
\affiliation{Institute of Cosmology and Gravitation, 
University of Portsmouth, Portsmouth PO1 2EG, Britain}
\affiliation{Institut f\"ur Theoretische Physik,
             Universit\"at Heidelberg,
             Philosophenweg 16,
             D-69120 Heidelberg,
             Germany}
\begin{abstract}

We study Randall-Sundrum brane models from the viewpoint of condensed 
matter/quantum optics. 
Following the idea of analog gravity we obtain effective
metrics in fluid and Bose-Einstein-condensate systems
mimicking those in the brane framework.
We find that the effect of warp factors in the bulk geometry translates
into finiteness of the analog systems.
As an illustration of this identification
we give a new interpretation of the peculiar behaviour
of the critical temperature for the splitting of thick
branes in warped spacetimes.

\end{abstract}
\maketitle
\def\half{{1\over 2}}
\def\L{{\mathcal L}}
\def\S{{\mathcal S}}
\def\d{{\mathrm{d}}}
\def\x{{\mathbf x}}
\def\v{{\mathbf v}}
\def\im{{\rm i}}
\def\etal{{\emph{et al\/}}}
\def\det{{\mathrm{det}}}
\def\tr{{\mathrm{tr}}}
\def\ie{{\emph{i.e.}}}
\def\bnabla{\mbox{\boldmath$\nabla$}}
\def\Box{\kern0.5pt{\lower0.1pt\vbox{\hrule height.5pt width 6.8pt
    \hbox{\vrule width.5pt height6pt \kern6pt \vrule width.3pt}
    \hrule height.3pt width 6.8pt} }\kern1.5pt}
\def\HRULE{{\bigskip\hrule\bigskip}}

\noindent
\section{Introduction}
Reasoning by analogy has been frequently the (hidden) seed 
behind the generation of new ideas in physics. 
Here we are interested in the existing 
analogies between gravitational and condensed-matter/quantum-optics realms
(see \cite{Workshop} for a review). They offer a bridge for interchanging 
conceptualizations of phenomena between strongly developed 
areas. In this letter we want to stress the potential
richness of this interrelation in connection with braneworld
gravities.

In the last years one of the gravitational configurations
that have attracted more attention have been the Randall-Sundrum (RS)
brane model with all its variants \cite{RS1,RS2}.
Essentially, the geometry of these models consist on a (4+1)-dimensional
asymptotically anti-de Sitter (adS) bulk separated into two symmetrical
parts through the time development of a 3-dimensional domain wall.
When leaving the domain-wall the spacetime becomes progressively warped
allowing the existence of a continuum of normalizable Kaluza-Klein
modes. The gravitational zero mode reproduces the standard Newton's law
on the brane, while the rest introduce short distance deviations
form the square law \cite{RS2,garriga}.

The metrics associated with thick brane generalizations of the 
RS one-brane model \cite{RS2} can be written as   
\begin{equation}
ds^2 = d\eta^2 +e^{2A(\eta)}[-dt^2+dy^2+dz^2+dw^2].
\end{equation}
or in conformal coordinates as
\begin{equation}
ds^2 = W^2(x)[dx^2-dt^2+dy^2+dz^2+dw^2],
\label{conformal-metric}
\end{equation}
with $x\equiv\int_0^{\eta} e^{-A(\eta)}d\eta$ 
accounting for the extra-dimension.
For instance, the (thin) RS case corresponds to
$A(\eta)=-|\eta|/l$ or $W(x)=1/(1+|x|/l)$, where $l$ stands for
the length scale of the warping which in $n$ dimensions is related 
with the bulk cosmological
constant by $\Lambda=-(n-2)(n-1)/2 l^2$. 
In this notation a thick domain wall in an asymptotically adS
space will correspond to different smoothings of the previous function
with the same asymptotic decay 
$W(x\rightarrow \pm\infty)\sim l/|x|$.

In what follows we will explain how these metrics (more properly their
4-dimensional counterparts which follows from eliminating one of the 
spatial coordinates on the brane, for instance the $w$ coordinate) can be
easily reproduced in an irrotational barotropic fluid and, with a bit
more complication, in a Bose-Einstein condensate. 
On the one side,
the existence of a brane in spacetime translates into the existence
of a domain wall in the fluid. Many of the concepts appearing
in surface critical phenomena, such as partial or complete wetting, 
will have their own equivalence in braneworld physics. 
These phenomena will play an important role in understanding
the collision and splitting of branes invoked in
some braneworld models \cite{turok}.
On the other side, we will see that the adS nature of the bulk translates
into finiteness of the condensed matter/quantum optics system behind the
analogy. 
Analyzing the critical features of the splitting of thick branes 
in an asymptotically
adS bulk one of the authors has found \cite{Cam:2002} that
the critical temperature for complete wetting is lower than that
appearing for asymptotically flat spacetimes. From the analog
model point of view, this modification of the critical temperature
can be seen as a finite-size effect. 

\noindent
\section{Fluid analog}
An irrotational barotropic inviscid fluid can be described 
by the equation of continuity 
\begin{equation}
\partial_t \rho + \bnabla\cdot(\rho \; \v) = 0,
\label{E-continuity}
\end{equation}
and Euler's equation
\begin{equation}
\rho {d \v \over dt} \equiv 
\rho \left[ \partial_t \v + (\v \cdot \bnabla) \v \right] =
- \bnabla p(\rho) - \rho \bnabla\Phi.
\label{E-euler}
\end{equation}
Here $\v=\bnabla \phi$ is the irrotational velocity of the fluid, 
$p=p(\rho)$ is the pressure, that only depends 
on the density $\rho$, and $\Phi$ represents the addition of all
possible external potential acting on the fluid.  It has been proved
that acoustic waves on a background fluid flow configuration (indicated by a
$0$ subscript) behave like a scalar field in a curved background
with the following 4-dimensional effective metric \cite{unruh,visser}
\begin{equation}
g_{\mu\nu} \equiv 
{\rho_0 \over  c_0}
\left[ \matrix{-(c_0^2-v_0^2)&\vdots&-v_0^j\cr
               \cdots\cdots\cdots\cdots&\cdot&\cdots\cdots\cr
               -v_0^i&\vdots&\delta_{ij}\cr } 
\right].               
\label{effective-metric}
\end{equation}
Here the velocity of sound in the fluid is defined as 
$c_0^2 \equiv (d p/ d\rho)|_0$.
We can easily see that in order to reproduce a metric of type
(\ref{conformal-metric}) we have to setup the velocity of
the flow equal to zero ($v_0^i=0$), and to assume a constant
velocity of sound; (absorbing this constant into a coordinate
redefinition we can set $c_0=1$). The constancy of the velocity
of sound implies that we have to deal with fluids with an
equation of state $p=c_0^2\rho$.

With these requirements the effective metric is obviously  
conformal to the Minkowski metric $g_{\mu\nu} = \rho_0 \eta_{\mu\nu}$.
To reproduce the conformal factor of a particular brane model one have only
to setup the fluid with a density profile of the form
\begin{equation}  
\rho_0(x)=W^2(x),
\label{density-profile}
\end{equation}
and this can be achieved by taking an external potential 
\begin{equation} 
\Phi=-2\ln W(x).
\label{external-potential}
\end{equation}
For instance, the original RS geometry corresponds to
\begin{equation}  
\rho_0(x)={1 \over (1+|x|/l)^2}; \;\;\;\; \Phi=2\ln(1+|x|/l).
\end{equation}

Given a spacetime geometry it is generically impossible to
reproduce it within a fluid analog, even taking advance of 
the freedom of coordinate choices (a generic 4-dimensional geometry 
has 6 degrees of freedom and a fluid analog has only two
and not completely independent).
What we have just seen is that the particularly simple way in which 
we can write the geometries of Randall-Sundrum type allows us to
build analog models for them.

Sonic waves in a background fluid with the above density profile
propagate in the same way as a (scalar) field would do in the RS
spacetime. 
The computation of geodesics on this geometry 
\cite{MucVisVol:2000}
shows that every free-falling test particle is repelled away from the
brane unless its trajectory is set initially on the brane itself.  A
repelled test particle reaches the adS horizon in finite
proper time. In the acoustic analog, this proper time corresponds
to the relative perception of time in the sonic world. 
A laboratory observer has a different perception of time based on
its underlying Minkowski spacetime. For him the proper time corresponds
to the coordinate time $t$ and the test particle would reach
the horizon when $t \rightarrow \infty$. Although from the 
sonic world point of view the geometry might be extended beyond
the horizon the reality of the system in the laboratory tell us 
that this extension is not actually in place. We are dealing with 
a genuinely incomplete effective geometry.

The behaviour of acoustic waves in these effective geometries captures
essential features of the propagation of gravitons in
spacetime. Basically, this is so because the background and the
acoustic waves are made from the same substance. Therefore, the
propagation of acoustic waves in these background geometries can be
understood from the braneworld point of view as a combination of zero
and massive Kaluza-Klein modes \cite{RS2}.  The zero mode will
correspond to wave packets localized on the brane and propagating
along the brane.  In realistic situations, in which the brane will
have some thickness, one expects that one can localize wave packets to
move along the brane if only one sets them up initially to have small
momenta in the transverse direction. If we associate a wavelength to
the transverse momentum $\lambda_t$, then this wavelength will have to
be much larger than the brane thickness, $\lambda_t \gg T$ for
localization to occurs. It would be interesting to analyze (and
possibly observe) this localization behaviour within the analog models
constructed.

Before finishing this section let us just anticipate that Eq. 6, apart
from being the key to construct the analog model, will allow an
stimulating identification between the effects of warp factors in
the bulk geometry and finiteness (finite-size) effects on condensed
matter systems.

\noindent
\section{Bose-Einstein analog}
Let us analyze now how would it be the construction requirements for
braneworld analog models within Bose-Einstein condensates.  These
systems are very promising for building analog models of black holes
\cite{garay}.

An appropriate starting point for describing Bose-Einstein condensates 
in dilute systems is the Gross--Pitaevskii equation 
\cite{dalfovo}
\[
\label{E:tdlg}
 \im \hbar \; \frac{\partial }{\partial t} \psi(t,\x)= \left (
 - {\hbar^2\over2m}\nabla^2 
 + V_{\rm ext}
 + \lambda \; \left| \psi \right|^2 \right) \psi(t,\x).
\label{eq:GP}
\]
In this approximation the interaction between the bosons in the 
condensate is modeled by a unique parameter 
$\lambda = (4\pi  \hbar^2 / m)\;a$, where $a$ is the s-wave scattering length.

By using the Madelung representation 
for the wave function
$\psi(t,\x) = \sqrt{\rho(t,\x)/ m} \; \exp[-i \theta(t,\x) / \hbar]$,
and defining $\v \equiv \bnabla \theta /m$
we can rewrite the (complex) Gross--Pitaevskii equation 
as a continuity equation and an Euler equation as in the fluid case.
In doing this we can associate to the condensate an
equation of state of the form $p=(\lambda/2m^2)\rho^2$
and consider the potential $\Phi$ in (\ref{E-euler}) to consist on 
two parts, $\Phi=V_{\rm ext}/m+V_{\rm Q}/m$,
the external potential seen from each boson
and the so-called quantum potential 
\begin{equation}
V_{\rm Q}(\rho) \equiv  - {\hbar^2 \over 2 m} 
\left( {\nabla^2 \sqrt\rho\over\sqrt\rho} \right).
\label{quantum-potential}
\end{equation}

Now, adopting the hydrodynamic approximation, which amount to
neglecting the quantum potential (we will comment later on some implications
of this approximation), one can proceed with the same steps
that in the fluid case to obtain an effective metric. 
Indeed, it is the same we have obtained before, but now for the sonic 
perturbations in the condensate \cite{garay,barcelo}. 
Again, in order that the 
metric (\ref{effective-metric}) reproduce braneworld metrics 
(\ref{conformal-metric}) we only need to have a constant 
sound speed $c_0=1$
and an specific density profile $\rho_0(x)=W^2(x)$.
However, the constancy of the sound velocity cannot be achived 
within a standard Bose-Einstein condensate.

A Bose-Einstein condensate can be seen as a particular type of fluid (in the
approximation we are dealing with) and so has a particular equation of
state.  Therefore, the speed of sound depends in a particular way on
the density
\begin{equation}
c_0^2={dp \over d \rho}\bigg|_0 = {\lambda \over m} \rho_0.
\end{equation}
Then, a profile in $x$ for the density requires for consistency a
counteracting profile for $\lambda$ or for $m$ if we want to keep
$c_0=$constant. This could be difficult to achieve in a laboratory but
can be envisaged from a theoretical point of view.  In condensed
matter physics one is used to encounter effective mass calculations
such as those for electrons immersed in a band structure (see, for
example, \cite{ziman}). They have been discussed also for excitons
(electron-hole pairs) in semiconductors.  By doping inhomogeneously
the semiconductor one could give place to a point dependent effective
mass. Concerning $\lambda$, or what is equivalent the scattering
length $a$, there is already many experiments in which the scattering
length is changed by using the properties of a Feshbach resonance
\cite{feshbach}. Essentially, in the surroundings of a Feshbach
resonance the value of the scattering length can be tuned by the
application of an external magnetic field. In current experiments this
magnetic field is homogeneous over the entire Bose-Einstein condensate
and the scattering length is changed only with time. One can imagine a
situation in which this modifications occur in an inhomogeneous way
without disrupting the condensate itself. From a more general
perspective, one can also envisage a situation in which the bosons
are immersed in a inhomogeneously doped medium resulting in
point dependent interactions.

In principle, given a brane geometry (that is, a warp factor $W$),
one can build its corresponding analog model by taking the suitable
density profile. We only need that the density profile vary smoothly
in distances of the order of the healing length defined as
$\xi\equiv \sqrt{2}\hbar/mc$. However, this is not possible in the
thin brane limit. In this situation the quantum potential
(\ref{quantum-potential}) cannot be neglected on the brane and thus
the hydrodynamic approximation no longer applies there. The quantum
potential will give a finite thickness to any physical realization of
a brane. Following the discussion on low-momentum approximation in
\cite{barcelo}, we can see that for waves with wavelengths greater
than the healing length, the effective geometrical description is
still in place, but with a different speed of sound. The new speed of
sound acquires additional (and non-homogeneous in the $x$ direction)
contributions coming from the quantum potential.  These contributions
are of the order of
\begin{equation} 
c_0{\rho \over m} {\xi \over T}.
\end{equation}
Therefore, strictly speaking with the previous procedure one can only
construct analog models for branes with $T > \xi$.  The effective
geometrical (low-energy) notions for the propagation of long wavelength
perturbations will show up in a different way in the confined region
if $T \sim \xi$.  This is the analog to what is expected to happen in
proper spacetime brane models. In order to understand the physics of
the confined region itself one might need a full-fledged string/M
theory.

Another important issue on building analog configurations with BEC is
their classical stability. At least for branes with $T > \xi$,
stability is guaranteed as in this approximation one does not have to
consider the quantum potential and so one is dealing just with a
condensate at rest inside a potential well.

\noindent
\section{Branes with constant curvature}
Up to now we have discussed how to reproduce a Minkowski brane
with an analog model. However, it is also possible to reproduce
other types of brane configurations.  

Let us consider now a 4-dimensional adS bulk with metric
\begin{equation}
ds^2 =  \Omega^2[dx^2-dt^2+dy^2+dz^2], 
\label{conformal-metric-4}
\end{equation}
where  $\Omega=1/(1+x/l)$ and
$x \in (-l,+\infty)$.
Starting from this spacetime we can built different types of $\mathbb{Z}_2$ 
(thin) braneworld geometries. If we take $\sigma$ to be 
the bare brane tension, the condition
\begin{equation}
{1 \over l}={1 \over 4 }\sigma \kappa^2,  
\end{equation}
where $\kappa^2$ is the gravitational coupling constant for the bulk, 
allows the existence of the Minkowski brane 
we have been analyzing. 
The $\mathbb{Z}_2$ symmetric spacetime that results from inserting 
a Minkowski brane into the adS bulk is obtained by simply changing $x$ 
in the metric (\ref{conformal-metric-4}) by $|x|$.
If the previous condition is not satisfied, we can introduce 
an additional length parameter $L$ as
\begin{equation}
\pm {1 \over L^2}=-{1 \over l^2}+{1 \over 16}\sigma^2 \kappa^4.  
\end{equation}

The negative sign case allows the existence of an adS
brane embbeded into the adS bulk. The metric is given by \cite{dewolfe}
\[
ds^2 = d {\bar x}^2 + 
F^2({\bar x})
\left[ d{\bar y}^2+ e^{-2{\bar y}/L} 
(-d{\bar t}^2+d{\bar z}^2) \right].
\]
Here $F({\bar x})=(l/L)\cosh[({\bar x}_h-|{\bar x}|)/l]$
with ${\bar x}_h$ an integration constant.
In the coordinates (\ref{conformal-metric-4}) an adS
brane will be located at
\begin{equation}
{y \over (l+x)}=\sinh\left({{\bar x}_h \over l}\right)\equiv s.
\end{equation}
It is worth noticing that $L$ does not appear in this expression.
This is because, from the bulk point of view, adS branes 
with different scales $L$ correspond only to different re-scalings of 
the coordinates in the brane. 

Let us define a function $\alpha_s(y)=(y/s)-l$.
Then, the metric associated to an adS brane
can also be written as in (\ref{conformal-metric-4}) but
substituting the conformal factor $\Omega^2$ by 
\[
W^2(x,y)\equiv \Theta(x-\alpha_s(y)) \Omega^2(x)+
\Theta(\alpha_s(y)-x)\Omega^2
\left(x_s^*(x,y)\right),
\]
where $x_s^*(x,y)$ implements the mirror symmetry across the brane
and is defined by
\begin{equation}
x_s^*(x,y)={1-s^2 \over 1+s^2}\;x + {2s^2 \over 1+s^2}
\;\alpha_s(y).
\end{equation}
To build the corresponding analog model one will
need a density profile and external potential characterized 
respectively by (\ref{density-profile}) and (\ref{external-potential}).

The positive sign case leads us to a de Sitter (dS)
brane model with metric \cite{dewolfe}
\[
ds^2 = d {\bar x}^2 + G^2({\bar x})
\left[ -d{\bar t}^2+ e^{2{\bar t} /L}
(d{\bar y}^2+d{\bar z}^2) \right].
\]
with $G({\bar x})=(l/L)\sinh[({\bar x}_h-|{\bar x}|)/l]$.
In the coordinates (\ref{conformal-metric-4}) the dS
braneworld will be located at
\begin{equation}
{t \over (l+x)}=\cosh\left({{\bar x}_h \over l}\right)\equiv c.
\end{equation}
As before, the warp factor corresponding to the dS brane can be written as
\[
W^2(x,t)\equiv \Theta(x-\alpha_c(t)) \Omega^2(x)+
\Theta(\alpha_c(t)-x)\Omega^2
\left(x_c^*(x,t)\right).
\]
However, in this case we cannot simply identify the corresponding
density profile to build an analog model of a dS brane.
By looking at the previous formula we see that this profile will
depend on time through $\alpha_c(t)$ and, therefore, the continuity
equation for the fluid will not be satisfied for every $x < \alpha_c(t)$.
In the coordinates we have analyzed it does not seem physically 
possible to build this analog model. 
By choosing a different coordinate system and considering more 
complicated analog models (see \cite{barcelo}) this problem might by 
alleviated.

\noindent
\section{Surface critical phenomena and finiteness effects} 
Phase transitions and critical phenomena play 
an important role in many areas of physics.
In the early universe, objects, such as domain walls, strings, and
monopoles are naturally produced in phase transitions.
In the context of brane cosmology the importance of 
critical phenomena appears at an even more fundamental level.
It has been suggested that some of the problems in standard
cosmology can be solved if our universe is the result of the splitting
and later collision of branes \cite{turok}.
Models of brane inflation can be generically constructed from the
collision and subsequent annihilation of branes and anti-branes, being
the inter-brane separation the degree of freedom associated with the
inflaton field \cite{DvaTye:1998}.
Brane gases has also been speculated to play a crucial role in
understanding the initial singularity and the origin of the spacetime
dimensions \cite{AleBraEas:2000}.

{}From condensed matter we know that the existence of surfaces in 
a bulk modifies and enriches the phenomenology of critical 
phenomena.
An interesting and illustrative situation is that of semi-infinite 
systems which
undergo a first-order bulk phase transition.
As the system is brought near the critical temperature of the
transition a disordered phase appears between the free surface
and the ordered bulk phase \cite{Lip:1982}.
The disordered phase could wet the free surface
partially, forming droplets attached to the surface, or completely, 
forming a layer of certain thickness. 
This type of critical phenomena presents a surprising effect:
the order parameters of the surface change continuously with 
the temperature while the order parameter associated with the
bulk varies abruptally at the critical temperature.
Essentially, this is a consequence of the coexistence of different 
types of vacuum states as the phase transition is reached.
Wetting of surfaces is a critical phenomenon that has been 
observed in many condensed-matter systems \cite{deG:1985}.
It has been found to appear also at the deconfined phase of SU(3) 
Yang-Mills theories \cite{wet_QCD} and supersymmetric QCD 
\cite{wet_SQCD}.

In a realistic realization of the RS brane 
scenario one would expect the bulk to be filled with a variety
of moduli and gauge fields. 
If the dynamics of some of these fields is driven by a phase 
transition the behavior of the embbeded brane could be altered
in different physically interesting manners.
Recently, it has been shown that the phenomenon of complete wetting 
also occurs for thick Minkowski branes embbeded in asymptotically adS 
spacetimes \cite{Cam:2002}.
A remarkable aspect in this case is the existence of 
a new effective critical temperature due to the presence 
of an effective negative bulk cosmological constant 
\cite{Cam:2002}. 
The phenomenon can be generically described by a phase 
transition supporting different discrete ordered phases 
which coexist with a disordered phase at a certain 
critical temperature $T=T_c$. 
As this temperature is approached an interface interpolating 
two different ordered bulk phases breaks into two separated 
interfaces, and a layer of disordered phase appears between them.
Contrary to what happens in asymptotically flat spacetimes, 
the width of this layer does not diverge for the temperature of vacuum
coexistence, $T_c$, but for a smaller effective critical 
temperature $T_\ast$.
This novel critical effect is directly linked with the global
geometry of the spacetime and can be easily understood
kinematically.
Since geodesics in RS backgrounds pull matter apart
from the brane \cite{MucVisVol:2000},   
once the splitting of branes is produced their mutual
repulsion makes the growth of the wetting layer to be enhanced.

{}From the analog-gravity perspective, this modification of the critical
temperature can be understood as a type of finite-size effect.
We have seen that the geometric warp factor translates into a density
profile that diminish progressively (vanishing asymptotically)
in the spacial direction that represents the extra dimension.
The decay of the profile is such that its integral over this dimension 
is finite. This means that, in what concerns the extra dimension, a finite 
amount of fluid has been spread out over an infinite space.
In practice the system will finish somewhere; however, with the
appropriate decay one can stretch the system as much as desire.
The finite-size effects we are discussing here could be more properly
called finiteness effects.
Finite-size effects are ubiquitous in quantum optics and condensed matter.
They generically produce modifications of the critical temperatures
associated with the thermodynamic (infinite) limit. 
The modified critical temperatures are observed to scale with the size 
of the system (in our case, related with the bulk cosmological constant).
Here, we want to suggest that many of the finite-size effects studied
in these areas, could have an interpretation in the field of brane
physics. 
Also from this viewpoint, it would be interesting to see what
can be learned by studying condensed matter/quantum optics models
with different types of finite-size effects: different confinement 
potentials would lead to finiteness or proper finite sizes;  
periodic conditions would translate into compactification, etc.

\noindent
\section{Summary}
In this letter we have built analog models for branes in warped spacetimes.  
They have helped us to established different links
between the behavior of braneworlds, and condensed matter/quantum optics
systems. 
We have identified warp factors in geometry with finiteness in analog 
systems. 
In our view, the understanding of brane physics (such as the
nucleation of bubbles, the splitting and collision of branes, etc.) when
they are immersed in a warped spacetime, can get important insights
from the analysis of surface critical phenomena in finite-size systems.  
The analog models we have constructed can guide us in the task 
of making a faithful translation of these phenomena into the 
realm of brane physics.

\noindent
\section{Acknowledgments}
CB is supported by the EC under contract HPMF-CT-2001-01203.
AC acknowledges the support of the University of Portsmouth,
the Univertit\"at Heidelberg and the Alexander von Humboldt
Foundation.




\end{document}